# Evidence for Mature Bulges and an Inside-out Quenching Phase 3 Billion Years After the Big Bang

**Authors:** S. Tacchella[1]*, C. M. Carollo[1]*, A. Renzini[2], N. M. Förster Schreiber[3], P. Lang[3], S. Wuyts[3], G. Cresci[4], A. Dekel[5], R. Genzel[3,6,7], S. J. Lilly[1], C. Mancini[2], S. Newman[6], M. Onodera[1], A. Shapley[8], L. Tacconi[3], J. Woo[1], & G. Zamorani[9]

**Affiliations:**

[1] Department of Physics, Institute for Astronomy, ETH Zurich, CH-8093 Zurich, Switzerland

[2] Istituto Nazionale di Astrofisica (INAF) – Osservatorio Astronomico di Padova, vicolo dell Osservatorio 5, I-35122 Padova, Italy

[3] Max-Planck-Institut für extraterrestrische Physik, Giessenbachstr. 1, D-85748 Garching, Germany

[4] INAF – Osservatorio Astronomico di Arcetri, Largo Enrico Fermi 5, I-50125 Firenze, Italy

[5] Racah Institute of Physics, The Hebrew University, Jerusalem 91904, Israel

[6] Department of Astronomy, Campbell Hall, University of California, Berkeley, CA 94720, USA

[7] Department of Physics, Le Conte Hall, University of California, Berkeley, CA 94720, USA

[8] Department of Physics and Astronomy, University of California, Los Angeles, CA 90095-1547, USA

[9] INAF Osservatorio Astronomico di Bologna, Via Ranzani 1, I-40127 Bologna, Italy

* sandro.tacchella@phys.ethz.ch; marcella.carollo@phys.ethz.ch

**Most present-day galaxies with stellar masses $\geq 10^{11}$ solar masses show no ongoing star formation and are dense spheroids. Ten billion years ago, similarly massive galaxies were typically forming stars at rates of hundreds solar masses per year. It is debated how star formation ceased, on which timescales, and how this "quenching" relates to the emergence of dense spheroids. We measured stellar mass and star-formation rate surface density distributions in star-forming galaxies at redshift 2.2 with ~1 kiloparsec resolution. We find that, in the most massive galaxies, star formation is quenched from the inside out, on timescales less than 1 billion years in the inner regions, up to a few billion years in the outer disks. These galaxies sustain high star-formation activity at large radii, while hosting fully grown and already quenched bulges in their cores.**

At the epoch when star-formation activity peaks in the universe (redshift $z\sim 2$; *1*, *2*), massive galaxies typically lie on the so-called "star-forming main sequence". Their star-formation rates (SFRs) tightly correlate with the mass in stars (stellar mass M), reaching up to several hundred solar masses ($M_\odot$) per year and producing a characteristic specific SFR (sSFR=SFR/M) that declines only weakly with mass (e.g., *3, 4*). In contrast, at the present epoch, such massive galaxies are spheroids with old stellar populations, which reach central surface stellar densities well above $10^{10}$ $M_\odot$ kpc$^{-2}$ and host virtually no ongoing star formation. Although the most massive ellipticals at z=0 bear the clear signatures of a gas-poor formation process (*5, 6*), the more typical population, at a mass scale of M~$10^{11}$ $M_\odot$, consists of fast rotators (*7*) with disk-like isophotes (*8*), steep nuclear light profiles (*9*) and steep metallicity gradients (*10*): all features that indicate a gas-rich formation process.

The full cessation of star-formation activity in these typical massive galaxies (here referred to as the quenching process) is not well understood, nor is its relation with the emergence of their spheroidal morphologies. Several quenching mechanisms have been proposed. The so-called halo-quenching scenario predicts that circumgalactic gas is shock-heated to high temperatures and stops cooling in dark matter halos above a critical mass (~$10^{12}$ $M_\odot$; *11*). Morphological/gravitational quenching proposes that the growth of a central mass concentration (i.e., a massive bulge) stabilizes a gas disk against fragmentation (*12, 13*). Feedback from an accreting supermassive black hole either transfers radiative energy (*14*) to the surrounding gas, thereby suppressing gas accretion onto the galaxy, or kinetic energy and momentum (*15*), which causes the expulsion of gas from the galaxy.

Quenching must soon occur in the most massive, and thus formidably star-forming, galaxies on the main sequence at z~2, to avoid dramatically overshooting the highest observed masses of z=0 galaxies (*16*). Yet no general consensus has emerged on which of the above mentioned processes is primarily responsible for halting this star formation as early as a few billion years after the Big Bang. Determining the distributions of the stellar mass and SFR densities within individual z~2 galaxies at high spatial resolution is central to resolving these issues. Together these distributions reveal how stellar mass builds up and SFR is progressively switched off inside these high-z galaxies which, given their high masses, will have to evolve into "red and dead" systems by z=0.

We measured such quantities for a sample of 22 star-forming galaxies at a median redshift of 2.2 (*17* and Section S1). The sample spans a wide range in stellar mass M~$4 \cdot 10^9$ to $5 \cdot 10^{11}$ $M_\odot$ and SFR~20 to 300 $M_\odot$ yr$^{-1}$ and broadly traces the main sequence at these redshifts.

The five most massive galaxies lie slightly below the average main sequence, a point we explore in more detail in Section S1.

For all galaxies we obtained adaptive-optics SINFONI (Spectrograph for INtegral Field Observations in the Near Infrared) spectroscopy on the European Southern Observatory's Very Large Telescope, mapping the two-dimensional rest-frame Hα emission at ~1 kpc spatial resolution. These data reflect the gas ionized by young stars within individual galaxies, which allows us to construct spatially resolved distributions of ionized gas kinematics and SFR surface densities internal to the galaxy. We also obtained Hubble Space Telescope (HST) imaging in the J and H passbands (*17, 18*). At the redshifts of the sample, the J and H filters straddle the rest-frame Balmer/4000Å break. This spectral feature is strong in relatively old stars and therefore provides a robust estimate of the stellar mass already assembled in older stellar populations. Thus, at a similar kiloparsec resolution as the SINFONI SFR maps, the HST images provide maps of the stellar mass density that is stored in such older underlying populations. Visual inspection of the two-dimensional SFR distributions immediately reveals their notoriously irregular appearance, with bright clumps at large radii, in contrast with their centrally peaked and smooth stellar mass distributions (Figs. S4-S6).

The shapes of the average surface SFR density ($\Sigma_{SFR}$) profiles (Fig. 1, top panels) are very similar regardless of total mass and are well fitted by a Sérsic profile $\Sigma \propto \exp(-\beta \cdot r^{1/n})$, with the $n$~1 value typical of disk-like systems. In contrast, the surface stellar mass density ($\Sigma_M$) profiles become progressively more centrally concentrated with increasing total stellar mass. The Sérsic index of the $\Sigma_M$ profiles increases from $n$=1.0±0.2 in the low mass bin, to $n$=1.9±0.6 in the intermediate-mass bin, up to $n$=2.8±0.3 in the high mass bin (uncertainties indicate the 1σ scatter). Within each mass bin, the mean $\Sigma_M$ profiles are always more centrally concentrated than the SFR density profiles.

We then compared the $\Sigma_M$ profiles of our sample of z~2.2 galaxies with a mass-matched sample of local z=0 galaxies (Fig. 1, bottom panels; *19*). Consistent with the Sérsic fits, the low-mass z~2.2 galaxies have the same radial stellar mass profiles of late-type disks in the local universe. The z~2.2 galaxies in the most massive bin, however, have stellar mass profiles that overlap with those of z=0 early-type galaxies out to galactocentric distances of a few kiloparsecs, corresponding to typically ~2 effective radii. At these high stellar masses (~$10^{11}$ $M_\odot$), our sample of galaxies on the z~2.2 have therefore already saturated their central stellar mass densities to those of galaxies of similar mass at z=0, which are quenched systems with a bulge-dominated morphology. Thus, the bulge components of these massive galaxies are already fully in place while their hosts are still vigorously forming stars farther out in the surrounding (disk) regions (Figs. 1 & 2, center panels; see also *13*).

Specifically, the ratio between the star formation rate and stellar mass surface density profiles (i.e., $sSFR_r = \Sigma_{SFR}(r)/\Sigma_M(r)$) indicates suppression of the inner $sSFR_r$ at the highest masses and generally outward-increasing radial profiles of $sSFR_r$ (Fig. 1, center panels). By examining the surface stellar mass density within 1 kpc, $\Sigma_{M,1kpc}$, as a function of total stellar mass M, we see that the massive galaxies at z~2.2 in our sample have substantially suppressed star formation activity in their centers relative to lower-mass galaxies at the same epochs. We show in the supplementary materials (*17*) that this cannot be entirely due to dust effects. The $sSFR_{1kpc}$ values range from $sSFR_{1kpc}$ ~5 $Gyr^{-1}$ (corresponding to a mass doubling time of ~200 Myr at a galaxy stellar mass of $10^{10}$ $M_\odot$) down to negligible values of ~0.1 $Gyr^{-1}$ at $10^{11}$ $M_\odot$. The key conclusion is therefore that these galaxies sustain their high total SFRs at large radii, far from their central dense cores, whereas in such cores the $sSFR_r$ is about two orders of magnitude lower.

A further important consideration that emerges from the analysis of Fig. 2 is that our galaxies lie around the identical tight $\Sigma_{M,1kpc}$-M sequence that is traced by galaxies at z~0 (*20*). This implies that the increase in total M of individual galaxies along the main sequence must be accompanied by a synchronized increase in central $\Sigma_{M,1kpc}$, until the maximal central stellar densities of today's massive spheroids are reached at a stellar mass scale on the order of ~$10^{11}$ $M_\odot$. Below this galaxy mass scale, dense stellar bulges are therefore built concurrently with the outer galactic regions. At z=0, the global relation curves, because the quenched galaxies have a shallower slope than the star-forming galaxies, and a clear "ridge" emerges (Fig. 2 ; *21*). We have too few galaxies to track this curvature at earlier times, but this trend would be consistent with z~2.2 galaxies slightly increasing their total M through declining star formation at large radii while maintaining their already quenched inner $\Sigma_{M,1kpc}$ values.

These results provide insight into the bulge formation process. The high stellar densities that are already present in their cores indicate that at least some massive star-forming galaxies at

z~2.2 have today's massive spheroids as their descendants. We are seeing, however, neither 'classical bulges' formed by dissipationless merging nor "pseudo-bulges" formed by the slow, secular evolution of a stellar disk. Such a dichotomy is often invoked to explain the structural variety observed in nearby galactic bulges (*22*; see, however, *23*). The high central stellar densities of the massive galaxies in our sample argue for a gas-rich, dissipative bulge formation process at even earlier epochs. This is consistent with theoretical predictions (*24*), that either mergers or violent disk instabilities in gas-rich galactic structures at high redshifts lead to a compaction phase of the gas component, which possibly even drags any pre-existing stellar component within the inner few kiloparsecs.

Furthermore, the suppressed central sSFRs in such massive systems argue for a quenching engine also at work. As this apparently acts from the inner galactic regions outward, it echoes findings recently reported for massive galaxy populations in a more recent era, at z~0.5 to 1.5, or ~1.5 to 5.5 Gyr later (*25*, *26*). Our results reveal that similar signatures are seen as early as z~2.2, implying that the same physical processes that lead to a phase of suppressed star formation from the inside out started acting on massive star-forming galaxies as early as ~3 Gyr after the Big Bang. We also estimate the time scale for an inside-out quenching wave to propagate across ~$10^{11}$ $M_\odot$ galactic bodies (Fig. 3). The estimate assumes that our galaxies keep forming stars with their observed radial profiles of surface SFR density until their $\Sigma_M(r)$ reaches the value observed in *z*=0 passive galaxies of similar stellar mass. This allows quenching timescales substantially less than 1 Gyr in the galaxy centers and roughly 3 Gyr in the outer disk/ring regions. These give rise to a stellar age gradient of dlog(age)/dlog(r)~-0.5 dex per radial decade (see *17*). This predicted age gradient in the stellar population implies a negative color gradient in passive z~1 to 2 spheroids, which is found in several studies (*27, 28*); with flat metallicity gradients, the inferred average age gradients range between about -0.1 and -0.4 dex per radial decade. A contribution to the color gradient from either dust or metallicity effects would imply that such estimates are lower limits to such photometrically estimated stellar age. The galaxies will be fully quenched by z~1; subsequent passive evolution down to z=0 will produce quenched z=0 remnants with the dissipative properties of typical ~M* spheroids, such as disk-like isophotes and fast-rotating kinematics (Fig. 4).

Our results also provide insight into quenching mechanisms. Clearly, an external process such as a large-scale shutdown of gas supply caused by a hot halo or a low cosmological accretion rate may still contribute to inside-out quenching. The fact that we observe a phase of inside-out quenching in very massive galaxies at early epochs suggests, however, also a prominent role of an internal process operating from the inner galaxy regions. The most massive galaxies in our sample exhibit fast nuclear outflows, which may indeed signify that active galactic nuclei feedback is also a factor (*29*). By setting the condition for sustainment of star formation, the local stellar density within galaxies may also be acting as the internal process that regulates the rate at which SFR is locally suppressed (*30*). The current analysis cannot identify the direction of the causality between the presence of a high stellar mass density and the cessation of star formation, but it is clear from our data that such high stellar density is present when quenching starts. Our study therefore highlights either, or possibly both, of these two internal processes as key contributors to the downfall of the most massive and most star-forming galaxies at the peak of galaxy formation.

**Acknowledgments:**

ST thanks B. Trakhtenbrot for stimulating discussions. We thank the reviewers for their thorough review, their comments and suggestions. We acknowledge support by the Swiss National Science Foundation, grant #200020_140952. This research made use of NASA's Astrophysics Data System, the arXiv.org preprint server, the Python plotting library matplotlib, and astropy, a community-developed core Python package for Astronomy. AR, GC and GZ acknowledge support from a INAF 'PRIN-2012'. Based on observations made with the NASA/ESA Hubble Space Telescope, obtained at the Space Telescope Science Institute, which is operated by the

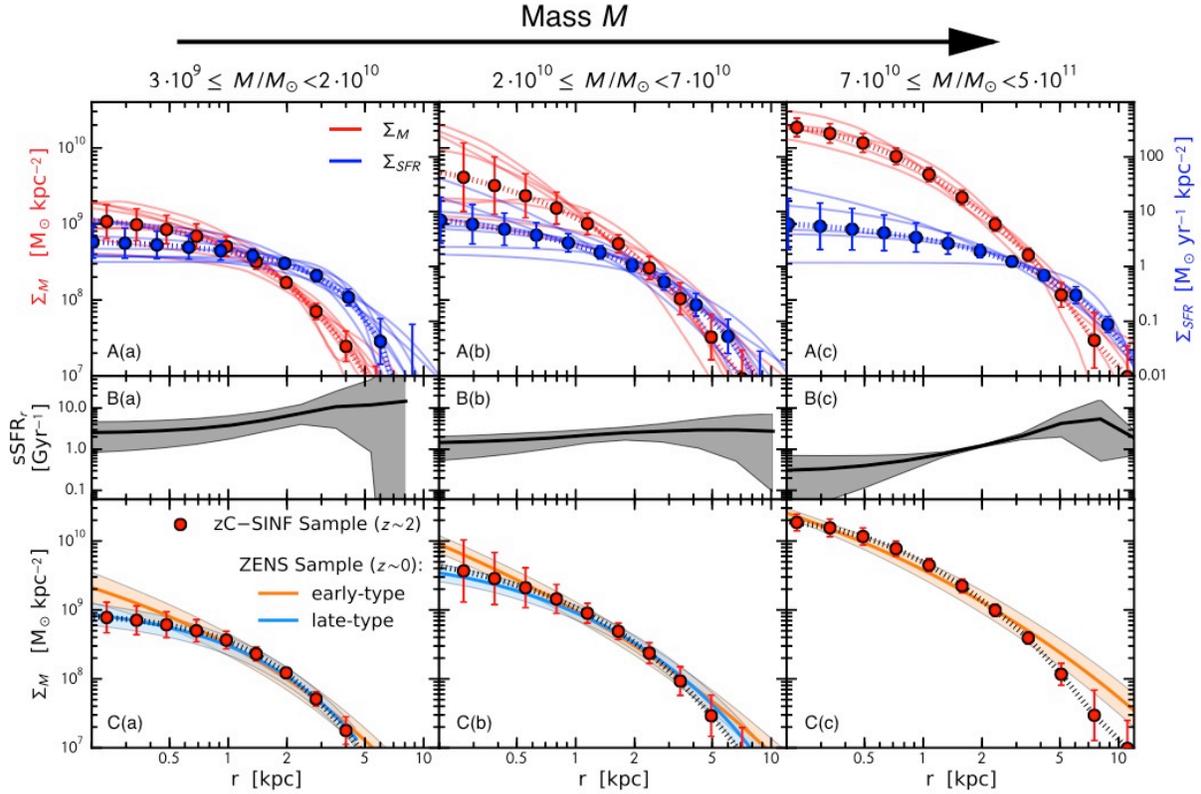

**Fig. 1. Stellar mass and star-formation rate surface density distributions.** The three columns of (a) to (c) show results for the three bins of stellar mass indicated at the top of each column containing 9, 8, and 5 galaxies, respectively. Upper row (A): The stellar surface mass density profiles (red, scale on the left vertical axis) and SFR surface density profiles (blue, scale on the right vertical axis) for our z~2.2 sample. Thin lines represent individual galaxies; the mean values are given by the solid circles (with error bars indicating the 1σ scatter). The derivations of these profiles are described in detail in the supplementary materials (*17*). Middle row (B): The mean sSFR as a function of radius r (sSFR$_r$; black line). In gray we show the 1σ scatter. Bottom row (C): The average surface stellar mass density profiles of the star-forming z~2.2 galaxies (red points with dashed line; error bars indicate the 1σ scatter) overplotted on the average profiles for the mass-matched samples of z=0 galaxies (colors indicate morphological types: orange for early-types, blue for late-types).

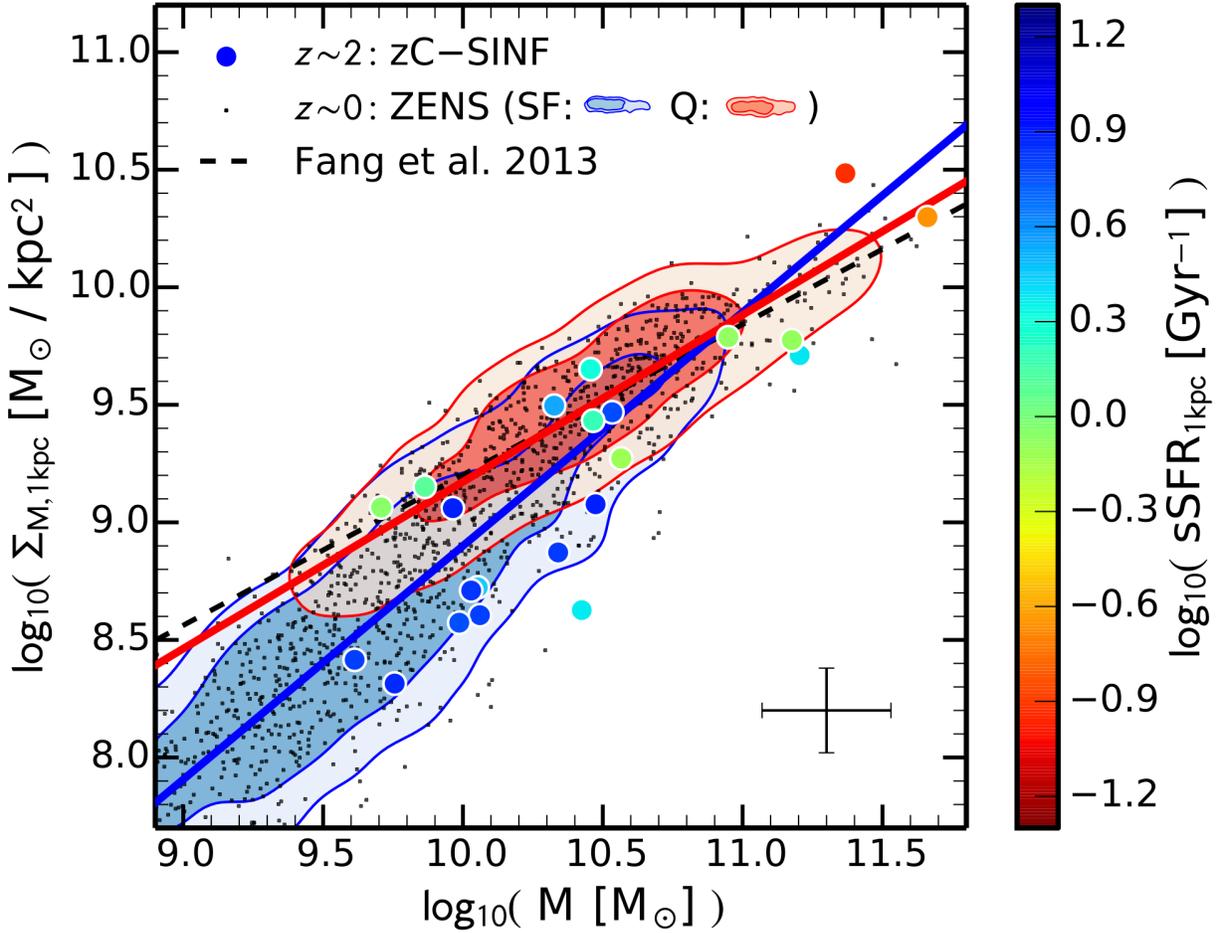

**Fig. 2. Central stellar mass density sequence.** We plot the stellar mass surface density within 1 kpc, $\Sigma_{M,1kpc}$, as a function of the total stellar mass M. The black points show the z~0 ZENS sample, and the blue and red contours their density on the $\Sigma_{M,1kpc}$-M plane split for star-forming (blue) and quiescent (red) galaxies. The correspondingly colored solid lines indicate the best fits to these z=0 star-forming ($\Sigma_{M,1kpc} \propto M^{\sim1.0}$) and quiescent ($\Sigma_{M,1kpc} \propto M^{\sim0.7}$) galaxies. The dashed black line shows the fit to the ridge of passive galaxies in the Sloan Digital Sky Survey z=0 sample of Fang et al. (*20*). The error bar at the bottom right indicate the systematic uncertainty in the derivation of $\Sigma_{M,1kpc}$ and M. The large points are the z~2.2 galaxies, color-coded according to specific star-formation rate within 1 kpc, sSFR$_{1kpc}$. The z~2.2 galaxies lie on the tight $\Sigma_{M,1kpc}$-M locus traced by the z=0 population. In contrast with total SFRs increasing with stellar mass along the main sequence, the z~2.2 galaxies have central sSFRs that strongly decrease with stellar mass.

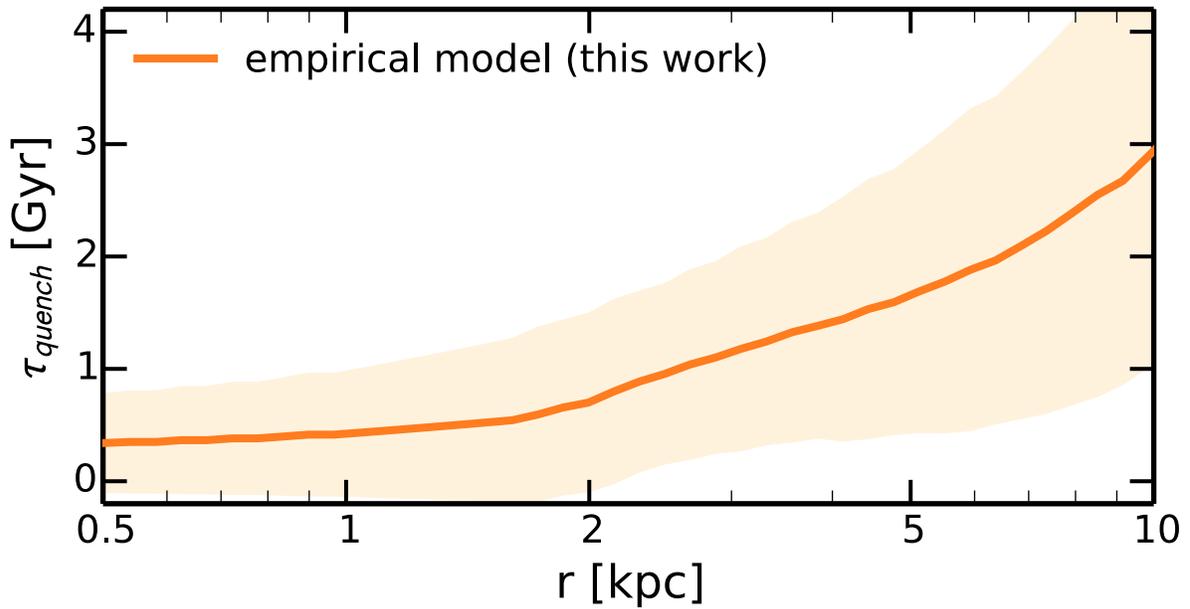

**Fig. 3. Outward progression of the quenching wave.** The quenching time $\tau_{quench}$ in star-forming $\sim 10^{11}$ M$_\odot$ galaxies at z~2.2 as a function of galactocentric distance. Such galaxies quench from inside out on time scales in the inner cores much shorter than 1 Gyr after observation, up to a few billion years in the galactic peripheries. The galaxies will be fully quenched by z~1. The solid orange line indicates the mean quenching time for all galaxies in the highest mass bin, whereas the orange-shaded region marks the 1σ scatter.

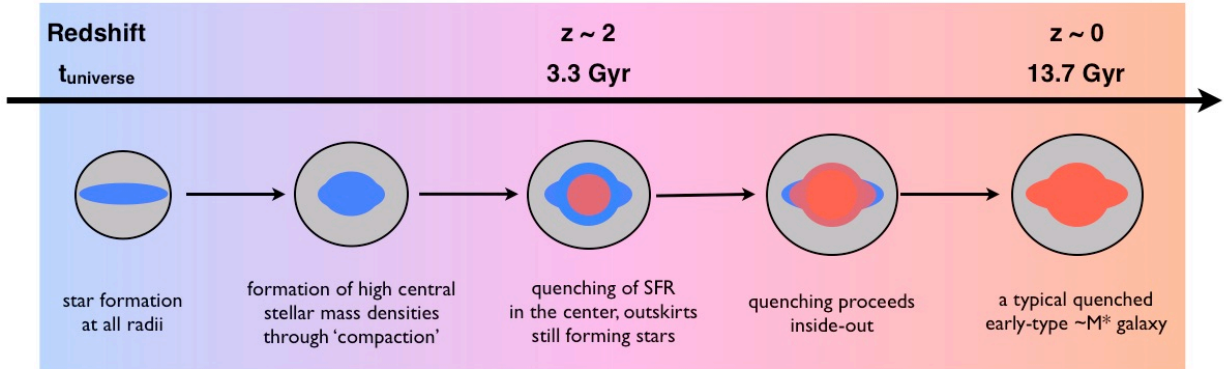

**Fig. 4. Proposed sketch of the evolution of massive galaxies.** Our results suggest a picture in which the total stellar mass and bulge mass grow synchronously in z~2 main sequence galaxies, and quenching is concurrent with their total masses and central densities approaching the highest values observed in massive spheroids in today's universe.

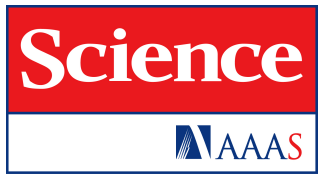

# Supplementary Materials for

Evidence for Mature Bulges and an Inside-out Quenching Phase 3 Billion Years After the Big Bang

S. Tacchella\*, C. M. Carollo\*, A. Renzini, N. M. Förster Schreiber, P. Lang, S. Wuyts, G. Cresci, A. Dekel, R. Genzel, S. J. Lilly, C. Mancini, S. Newman, M. Onodera, A. Shapley, L. Tacconi, J. Woo, & G. Zamorani

\*correspondence to: sandro.tacchella@phys.ethz.ch; marcella.carollo@phys.ethz.ch

**This file includes:**

Materials and Methods S1 to S3
Supplementary Text S4 to S5
Figs. S1 to S10
Tables S1
References (*31-58*)

In this Supporting Online Materials we provide additional details on the data sets and the data analysis regarding the article 'Evidence for Mature Bulges and Inside-out Quenching 3 Billion Years After the Big Bang'. As part of Materials and Methods, we first describe the galaxy sample and summarize basic measurements done on this data (Section S1). In Section S2 we introduce our z=0 comparison sample. In Section S3 we describe in detail the derivation of the stellar mass and SFR profiles and maps. In Section S4 we investigate the impact on our results of different radial boundaries for estimating parameters in the central regions, and of different dust models. In Section S5 we explain in detail the model that we have adopted for computing the quenching timescales discussed in the main text. Throughout, we use the following cosmological parameters: $H_0 = 69.3$ km s$^{-1}$ Mpc$^{-1}$, $\Omega_{\Lambda,0} = 0.71$, and $\Omega_{m,0} = 0.29$ (WMAP9, see *31*). For this cosmology, 1" corresponds to ≈8.4 kpc at z=2.2. Magnitudes are given in the AB photometric system. All sizes and radii presented in the paper are circularized, i.e. $r = r_a\sqrt{(b/a)}$, where $r_a$ is the size along the semi-major axis and (b/a) the minor/major axis ratio.

**Materials and Methods**

**S1 Our Sample of 22 Main Sequence Galaxies at z~2.2**

Our 22 galaxies are drawn from the Spectroscopic Imaging survey in the Near-infrared with SINFONI (SINS, see *32*) and zCOSMOS-SINFONI sample (zC-SINF, see *33*) which comprise together a total sample of 110 massive z~1-3 SFGs with seeing-limited SINFONI data. Out of this parent sample, we followed up 35 galaxies with SINFONI Adaptive Optics. Depending on the redshift of the sources, we used the K-band (z > 2) or H-band (z < 2) grating to map the main emission lines of interest (Hα and the [NII]λλ6548,6584 doublet). The nominal spectral resolution is R~2900 and 4500 for the H- and K-band, respectively, i.e. we are able to not only separate the [NII]λλ6548, 6584 doublet from Hα, but we resolve the gas motions within the galaxies well. This is a clear advantage to HST grism imaging, since we are able to map the internal kinematics with a ~1 kpc resolution and remove potential uncertainties from [NII] emission contribution to our Hα fluxes.

The observations and reduction of the complementary HST near-IR data for 29 out of the 35 SINFONI AO observed galaxies are described in Tacchella et al. (*18*) and Förster Schreiber et al. (*34*). All of these galaxies have at least one orbit with HST's Wide Field Camera 3 (WFC3) J-band and two with HST H-band. Beside the data reduction, Tacchella et al. (*18*) present all the basic measurements on the light such as the PSF-corrected J- and H-band surface brightness and observed J-H color profiles (i.e. rest-frame u-g color) of these 29 galaxies.

In this work, we focus on the 22 galaxies for which the light distribution is well described by a PSF-corrected Sérsic profile. In five cases, the presence of bright clumps complicate the Sérsic fits to the J- and H-band light distributions. Simple centrally concentrated model profiles do not adequately represent the extreme 'ring-like' light profiles of these objects (Q2343-BX389, Q2343-BX610, Q2346-BX482, ZC405501, and ZC406690; see (*18*)). At the SINFONI resolution, however, the Hα kinematics seem regular and consistent with disk rotation despite the clumpy appearance (*13*). In addition, the galaxies K20-ID7 and Q1623-BX502 are also excluded since their mass estimates are very uncertain, as indicated by > 1 order of magnitude difference between the estimate based on the observed J-H color and that based on the SED fit (see S3.6).

Our sample of 22 galaxies with their stellar population properties is listed in Table S1. The stellar mass, age, visual extinction ($A_V$), absolute and specific SFRs were obtained from evolutionary synthesis modeling of the optical to near-IR broadband spectral energy distributions (SEDs) supplemented with mid-IR 3-8 μm photometry when available. We define the galaxy's stellar mass as the mass obtained from the integration of the star formation history. All details of the SED modeling are presented in Förster Schreiber et al. (*32*) and Mancini et al. (*33*). Briefly, we use the Bruzual & Charlot models (*35*), a Chabrier initial mass function (IMF) (*36*), solar metallicity, the Calzetti et al. reddening law (*37*), and either constant or exponentially declining SFRs.

Figure S1 shows that the five most massive galaxies in our sample are on average 0.44 dex below the main sequence average (with 4 out of 5 less than 0.4 dex and one at 0.7 dex below), which itself has a 1-sigma scatter of ~0.3 dex, i.e. they are *marginally* 'sub main sequence' (see also *18*, *38*). It is legitimate to question whether this is related to having a quasi-quenched inner 'bulge', as we illustrate in the main text. In the case of our five galaxies, the sSFR in the inner 'bulge' and outer 'disk' regions are ~0.3 $Gyr^{-1}$ and ~5 $Gyr^{-1}$, respectively (see Figure 1). This is a difference in sSFR of the bulge relative to the disk regions of a factor of ~17. If the inner regions had the same sSFR as the surrounding disks, the galaxies would move up in SFR by ~0.4 dex, i.e., right on the average main sequence. A larger sample of 'sub main sequence' galaxies will be required to demonstrate whether, at constant stellar mass, the SFR 'distance' below the main sequence correlates with the amount of mass in the bulge component with suppressed sSFR. Our sample is too small to make this additional claim. It does however demonstrate, as we emphasize in this paper, that at least our sample of high masses galaxies at these high redshifts do host quenched bulges along with their actively star-forming disks. By the same token, galaxies that are still on or above the main sequence may have not started to quench in their central regions.

**S2 The z~0 Comparison Samples**

We use our Zurich Environmental Study (ZENS; *19*) as our main local comparison sample. The ZENS sample consists of 1600 galaxies within the narrow redshift range 0.0500 < z < 0.0585. In Cibinel et al. (*39*, *40*), a careful analysis of the galaxy structural and photometric measurements as well as morphological classifications are presented. The choice of adopting ZENS as our z=0 benchmark, rather than using catalogued data for a much larger sample such as the SDSS (*41*), is motivated by two main considerations. First, our start-to-end understanding of its data quality and limitations, its sample properties, and all assumptions behind the measured quantities; second, the fact that its manageable sample size has allowed us to inspect and cross-check, for each galaxy individually, all parameter measurements, thereby minimizing the impact of systematic errors in the data processing and resolving any inconsistency in such parameters.

We derived the stellar mass surface density profiles for the ZENS galaxies from B-I color profiles obtained from single Sérsic fits to B- and I-band photometry, following the same conceptual approach as with our z~2.2 galaxies (see S3 for a detailed description of the approach). Specifically, for each galaxy, we derived its mass profiles using the individual relation between its B-I color and its M/L ratio obtained considering its redshift and dust reddening (see *40*).

Finally, we explore whether the specific choice made for the fiducial z=0 comparison sample, and the different rest-frame colors used to model the stellar masses of the z=0 and z~2.2 galaxies, introduce systematic biases that might affect our main conclusions. Such biases might arise from *(i)* either biases in the specific sample relative to the global local population, *(ii)* or in the mass estimates based on different input rest-frame colors relative to using identical rest-frame colors. As shown in Fig. 2 through the direct comparison with the $\Sigma_{M,1kpc}$-M relation of (*21*), and also extensively discussed in (*39, 40*), there are no biases in either the mass estimates or the surface mass density (including central values) estimates for ZENS galaxies relative to the much larger SDSS population. Fig. S2 furthermore shows a very good agreement between the stellar masses obtained, for the SED model templates of our library, when using the rest-frame B-I color, as done for the ZENS galaxies, and the rest-frame u-g color, which is the input color that is provided by the J-H color at the redshift of our high-z sample. We therefore stress there are no spurious biases, that may affect our conclusions, that arise from having adopted the well-studied ZENS sample as our fiducial z=0 comparison benchmark.

**S3 Mass Profiles and Star-Formation Rate Profiles**

In this Section, we briefly describe how the color profiles for the z~2.2 galaxies were obtained from analytical fits to their surface brightness distributions; details are given in (*18*). A description of the derivation of the stellar mass surface and star-formation rate density profiles follows. In the last Subsection, we obtain upper limits on possible active galactic nuclei (AGN) contributions.

*S3.1 Stellar Population Modeling*

In this section, we describe the stellar population modeling to derive a relation between M/L and observed J-H color for our z~2.2 galaxies, i.e., the rest-frame u-g color. For the z=0 ZENS sample, a similar relation is used based on the B-I color. In Section S3.2 and S3.5 we use this relation to derive the mass distribution from the J-H color within galaxies. The strength of using the J- and H-band filters for z~2.2 galaxies is that J-H color minimizes extinction effects, optimally brackets the age-sensitive Balmer/4000Å-break, and the wide bandpasses maximize flux sensitivity.

It is well known that observed colors and stellar mass-to-light ratios (M/L) of stellar populations are correlated. We use this relation to construct stellar mass profiles (and M maps) from our color profiles (color maps) in order to investigate spatial variations in stellar mass on resolved scales of ~1 kpc.

We employ Bruzual & Charlot models (*35*) to generate the observed J-H colors of stellar populations. The colors are obtained from the redshifted model spectra using the total transmission curves of the WFC3 J- and H-bands. We use the quantity $\log_{10}(M)+0.4 \cdot H$, which is proportional to the $\log_{10}$ M/L and directly linked to our observed quantities. The correlation between $\log_{10}(M)+0.4 \cdot H$ and rest-frame optical color is displayed in Fig. S3 where evolutionary tracks in the observed J-H color − $\log_{10}(M)+0.4 \cdot H$ plane are shown for various star formation histories (SFHs), shown by different colors, and metallicities. We consider a variety of SFHs ranging from declining SFRs with e-folding timescales from 50 Myr to 3 Gyr, and constant star formation. The models are computed for a grid of ages from 10 Myr to the age of the universe at the given redshift, of metallicities (solar and 1/5 of solar), and of extinctions with $A_V$ from 0 to 3 mag (different line-style). The main effect of increasing extinction is to shift the model curves along a direction that is roughly parallel to the age tracks. The model curves occupy a fairly well defined locus in the observed J-H versus $\log_{10}(M)+0.4 \cdot H$ parameter space, reflecting the degeneracy between stellar ages, extinction, and SFH in these properties. We use this degeneracy to derive the mean effective $\log_{10}(M)+0.4 \cdot H$ for a given observed color, which, adding the observed H-band, yields a stellar mass estimate without requiring any knowledge about the actual SFH, age, or extinction. Other SFH families (e.g., delayed tau models, increasing tau models) occupy the same locus in these diagrams.

We compute the mean relationship by taking the mean value in $\log_{10}(M)+0.4 \cdot H$ over all model curves within bins of width 0.1 mag in J-H colors. The mean relationship for solar metallicity is plotted in left panel of Fig. S3 with gray-filled circles, with error bars corresponding to the standard deviation of all models within each color bin. The relationship derived for 1/5 solar metallicity is also shown for comparison with white-filled circles, but is very similar to the one for solar metallicity. The right panel of Fig. S3 shows how the mean relation changes with varying redshift. We have calculated the mean relation, at the redshift of each galaxy, accounting for different J-band filters (archival HST J-band imaging is done with the F125W filter, while our follow-up program uses the F110W filter).

*S3.2 Derivation of Color and Mass Profiles*

The center of each galaxy for the fitting was carefully chosen and uncertainties in the center position were propagated to the uncertainty estimation of the fits. We used GALFIT (*42*) to fit the light distribution with single Sérsic profiles (*43*). We identified adjacent objects from segmentation maps obtained from SExtractor (*44*). These were masked out or, when too bright and/or too close, so to generate overlapping isophotes, they were fit simultaneously to our target. We used the resulting single Sérsic profiles to construct the PSF-corrected J-H color profiles for all the galaxies, from which we derived the surface stellar mass density profiles using the relation described in S3.1 and the observed H-band luminosity profiles. The errors in the mass profiles are estimated from the errors in the J-H color, the error in the conversion from color to M/L, and the error on the H-band luminosity profile.

*S3.3 Derivation of SFR Profiles*

The SFR surface density profiles were obtained from the SINFONI Hα emission line maps that we present in Förster Schreiber et al. The Hα emission is very clumpy and irregular, i.e. a single 2d Sérsic fit did not converge to sensible results using GALFIT directly on the map. We therefore converted the 2d maps to a 1d profile by measuring the flux in elliptical apertures with the same ellipticity and center as obtained by the H-band GALFIT fits (using IRAF task ELLIPSE). The 1d profiles were then fitted taking into account each galaxy's PSF.

We assume that the measured narrow component of the Hα-line emission originates from star-forming regions, with no contribution from AGN or shock-ionized material (other sources should be negligible for actively star-forming galaxies; e.g., *45*). Since we do not have direct estimates of the dust attenuation applicable to Hα line emission (e.g., from measurements of the Balmer decrement), we use the best-fit extinction values $A_{V,SED}$ from the SED modeling (Table S1). The Calzetti law acts as an effective foreground screen of obscuring dust, and so the extinction correction at the wavelength of Hα is $L(H\alpha)=L_{obs}(H\alpha) \times 10^{0.33 A_{V,neb}}$, with $A_{V,neb}=A_{V,SED}/0.44$. We then converted this dust-corrected Hα luminosity to the SFR using: $\log(SFR(H\alpha)) = \log(L(H\alpha)) - 41.10$ (*46*), adding however a -0.23 dex to the right-hand term to take into account of the different IMFs used by reference (Salpeter IMF) and in our work (Chabrier IMF). The main uncertainty in this derivation is the assumption of spatially homogeneous foreground extinction towards stars and HII regions, which is not necessarily the case (see Section S4 for a detailed investigation).

*S3.4 Emission line contribution and AGN contamination*

The stellar mass estimates from the observed J- and H-band color (rest-frame u-g color) can be affected by strong emission lines from star-forming regions as well as AGN: [OII]3726,3729 contribute to the flux in the J-band, while Hβ and [OIII]4959,5007 contribute to the H-band. From the line strength of Hα (Hα-fluxes ranging from 2.9 – 30.3 x $10^{-17}$ erg s$^{-1}$ cm$^{-2}$) we can estimate the fluxes of the other lines. These line ratios are highly uncertain, but, if we use sensible ranges, i.e., for Hβ/Hα, [OIII]/Hα, and [OII]/Hα respectively 0.3-0.4, 0.15-2.0, and 0.15-2.0 (*47*), the correction are of order 16% and 12% to the H- and J-bands, respectively. These corrections compensate each other, so ultimately the impact on our mass estimates are expected to be <10%, i.e., well within the errors.

For the most massive galaxies (ZC400528, ZC400569, D3A-6004, and D3A-15504), we have measured nuclear outflows (*29, 48*): the spectra in their central regions exhibit a broad component in Hα, forbidden [NII] and [SII] line emission, with typical velocity FWHM ~ 1500 km/s, and intrinsic extent of 2-3 kpc. These properties are consistent with warm ionized gas outflows associated with Type 2 AGN, the presence of which is confirmed via independent diagnostics in half the galaxies. As mentioned above, the Hα line maps/profiles (i.e., the star formation distribution) is inferred from narrow component of the Hα-line emission, which is corrected for the contribution of broad components of winds (stellar or AGN). An additional uncertainty in the stellar mass estimates is the emission from the AGN itself, which can contribute significantly or even outshine the rest-UV to near-IR emission from the stellar populations of the host galaxy. We expect that, for Type 2 AGN, the broad band rest-optical colors and derived M/L are unlikely to be significant affected (e.g., *49, 50*). Specifically for our galaxies, Genzel et al. (*48*) and Mancini et al. (*33*) showed that the SEDs for all our objects are well represented by pure stellar templates. We therefore conclude that possible AGN contribution does not influence the estimated quantities that we use in this work.

*S3.5 Derivation of 2d Stellar Mass Density Maps*

Although in this work we chose to use the one-dimensional stellar mass profiles constructed from the Sérsic fits to the azimuthally-averaged, J- and H-band light profiles presented in (*18*), we also constructed the full two-dimensional J-H color maps from the J- and H-band images. An important step in the construction of these maps is the application of a Voronoi tessellation algorithm to locally enhance the signal to noise, especially in the outskirts of the galaxies. The color maps are converted into stellar mass (M) maps by using again the J-H color – $\log_{10}(M)+0.4 \cdot H$ relation discussed in Section S3.1. These M maps are shown in Figs. S4-S6 (2nd column). The errors (68 % confidence interval) in the M maps include the errors in the color maps, in the color – $\log_{10}(M)+0.4 \cdot H$ conversion, as well as in the H-band image. The uncertainty of the M maps in the inner parts of the galaxy amounts to a factor of ~2, while it increases towards the outskirts to about a factor of ~10 (about 10 kpc away from the center). As also noted by, e.g., (*51-53*), overall the M maps are smoother than the H-band light distributions and Hα-line emission maps. Star-forming clumps in the galaxies' outskirts are clearly visible in the Hα map, and sometimes even in the H-band images, but they are barely noticeable in the M maps.

There are pros and cons in using one (1d color profiles from Sérsic fits) or the other (2d mass maps) of the two approaches for estimating the spatially-resolved stellar mass densities within the z~2.2 galaxies. Relevant to our study, however, in the next Section we present a straight comparison of both approaches and show that both converge to very similar results.

*S3.6 Verification of Mass and SFR Density Profiles and 2d Maps*

In Fig. S7 we compare the total stellar mass obtained by integrating the M maps (left panel) and the surface mass density profiles (right panel) with the stellar masses obtained from the SED fitting to the 22 galaxies of our sample; the resulting ~45º slope tight correlation demonstrates the good agreement between the two estimates, validating the self-consistency of the derived 2d $M_\star$ maps. We note nevertheless that the SED fitting mass is systematically slightly higher (less than a factor 1.5) than the other estimate. The SED fitting masses are derived from the aperture-corrected photometry. On the other hand, the M maps are obtained after summing up all pixels with S/N>5, which leads to somewhat smaller effective apertures than for the integrated broad-band photometry. This effects fully explains the small offset, which we stress does not affect either our results or our discussion.

Figs. S4-S6 show the H-band light distributions ($1^{st}$ column), M maps ($2^{nd}$ column), mass profiles ($3^{rd}$ column), SFR maps ($4^{th}$ column), and SFR profiles ($5^{th}$ column, solid lines; these are fits to the 1d profiles obtained from the SFR maps, black points). In the $3^{rd}$ column, the red and orange lines are the 1-d stellar mass profiles obtained from the Sérsic fits to J- and H-band, respectively PSF-corrected and PSF-convolved. The overplotted black data points are obtained through azimuthally-averaging the M maps. For the vast majority of the galaxies, the surface mass density profiles obtained from the '1d' and '2d' approaches agree remarkably well within the errors, which gives strong confidence that there are no strong biases introduced when using one or the other of the two methods.

**Supplementary Text**

**S4 The central sSFR and Dust**

In our study we conclude that SFR is suppressed in the central regions; e.g., Fig. 2 uses the central sSFR within 1 kpc. A first question to ask is whether this main conclusion depends on the precise size of the aperture that we use to define the 'central' sSFR. The cenral panels of Fig. 1 already show that this is not the case. We also add in Fig. S8 the equivalent of Fig. 2, produced however using two other definitions of central aperture: the quarter of the half-mass radius $R_e$ and a 2 kpc radius. In the upper panel, we plot the central stellar mass density within 1 kpc as a function of mass with color-coding of the sSFR within these two additional apertures; in the lower panels, we show central stellar mass densities calculated within the same apertures as the sSFR. There is some reduction in dynamic range of the sSFR within the larger aperture(s), due to a larger inclusion of star-forming outskirts. Nevertheless, the trend that we discuss in the main text, i.e., that the most massive galaxies have reduced central star formation activity, remains valid also when using these alternative apertures.

A second issue to address is that our approach assumes that the shape SFR profile is identical to that of the Hα profile, with uniform scaling that accounts for dust attenuation. This may not be the case (see, e.g., *26, 52*). Centrally concentrated dust would cause the sSFR to appear spuriously low in the inner kpc. This would require however a total dust extinction in the central regions of the most massive galaxies of $A_{V,SED} \approx$ 3-5 mag, and ~3 mag of additional extinction in higher mass relative to lower mass galaxies. Measured $A_{V,SED}$ values for star-forming galaxies at these redshifts are in the 0-2 mag range (*26, 54, 55*), with a typical value of 0.8 mag.

Also, in addition to spatially inhomogeneous extinction, we explore the possibility of differential extinction between Hα and the underlying continuum. At 0.7<z<1.5, and for galaxies with $\log_{10}(M/M_\odot)>10.7$, Wuyts et al. (*26*) have suggested a somewhat different Hα dust correction than the Calzetti law with $A_{V,neb}=A_{V,SED}/0.44$ that we have adopted as our benchmark. Specifically, the Wuyts et al. calibration is: $A_{V,neb}=1.9\ A_{V,SED} + 0.15\ A_{V,SED}^2$, and is consistent with results from an independent Balmer decrement methodology by (*55*). We plot its implementation in Fig. S9 (which is a remake of Fig. 2). The sSFR is slightly reduced for all galaxies (<0.2 dex), the effect being stronger, about ~0.3 dex, in less massive galaxies. Overall, our main result however does not change.

We therefore conclude that dust extinction may contribute to mimic a suppression of SFR in their inner regions, but that most of the effect that we discuss in the text is due to a genuinely reduced central star formation activity.

**S5 Estimating the Quenching Timescale**

The probability that a galaxy quenches is a strong function of mass: local galaxies with a total stellar mass above ~$10^{10.85}$ $M_\odot$ are usually quenched in the local universe (e.g., *56*). In the analysis presented here, in addition to the total integrated stellar mass, we also know how the stellar mass is distributed within the galaxies. For galaxies which are already very massive at

z~2.2, this gives us a resolved constraint on the growth of its stellar mass in the sense that such galaxies cannot arbitrarily increase their stellar mass density at a given radius in order not to overshoot the stellar mass density observed in the similarly high-mass local galaxies.

We use this argument to constrain the time when the most massive galaxies in our sample quench their star formation at a given radius, by evolving their stellar mass density profiles over time by adding their integrated SFR profile to the z~2.2 stellar mass profiles, while assuming that the total SFR of the galaxies follows that of a typical main sequence galaxy. We then compare the evolved stellar mass density profiles at any given redshift to the stellar mass density profiles of similarly massive z=0 spheroids; specifically, once the stellar mass density reaches the z=0 value at a given radius, star formation is assumed to cease and this is taken to be the time of quenching at that radius.

An uncertainty in this approach is which mass is assumed for the z=0 descendants of our z~2.2 galaxies. Using abundance matching methods or simple continuity arguments (*56*, *57*), our lower mass galaxies ($10^9$-$10^{10}$ $M_\odot$) grow by a factor of ~30 and will end up in a wide range of z=0 descendants, depending on their environment. It is thus not possible to make a prediction for their quenching time scales. In contrast, the most massive galaxies in our sample (~$10^{11}$ $M_\odot$) will grow by at most a factor of ~4, depending on the contribution of merging (i.e., the environment): it is thus for these very massive galaxies only that we use our simple model to derive an indication of their local quenching time scales.

In Fig. S10, we show the result of our model for the mean stellar mass and SFR profile of the galaxies in our most massive bin. The integrated profiles amounts to M = 2.1·$10^{11}$ $M_\odot$ and SFR = 157 $M_\odot$ yr$^{-1}$ at z=2.2. The total mass increases by a factor of 2.6 down to z=0. The stellar mass within 2 kpc reaches the maximum given by the z=0 mass profile before the outer regions, i.e. the galaxy quenches first in the center and proceeds radially-outward towards later epochs. The estimated time scales range from >300 Myr after observations in the inner regions, up to 1-3 Gyrs in the galaxy outskirts, giving rise to an age gradient of dlog(age)/dlog(r)~-0.5 dex per radial decade (see Fig. 4). The galaxy is fully quenched by z~1.

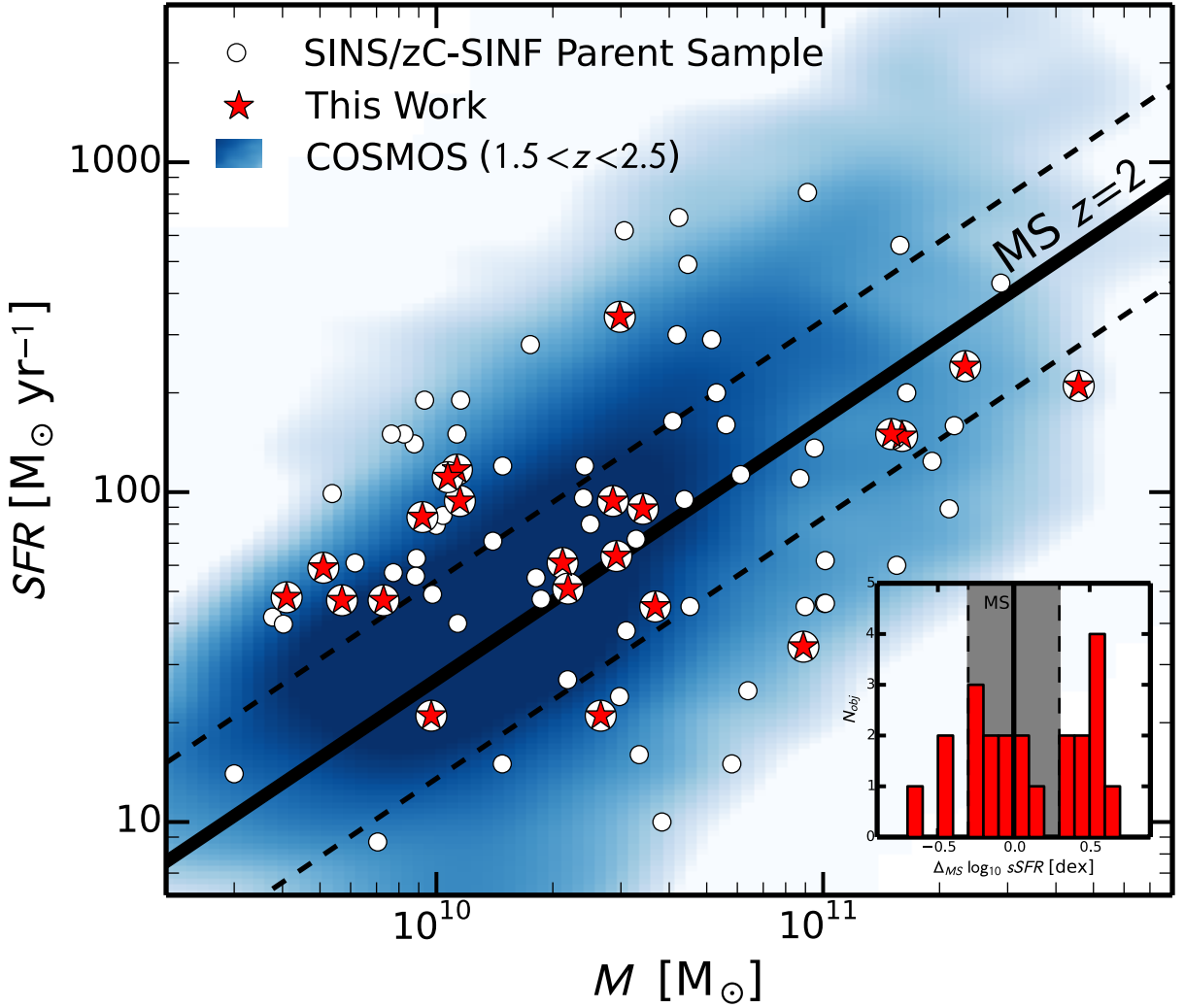

**Fig. S1. Location of our sample on the z~2 star-forming main sequence.** The black solid line shows the main sequence and dashed lines indicate its 0.3 dex scatter. The white circles show the SINS/zC-SINF parent sample, the red stars are our sample, and the blue shaded regions shows the distribution of ~6000 SFGs from the COSMOS survey (*58*) in the redshift range 1.5<z<2.5. The inset shows the distributions of the offsets in specific SFR from the main sequence (in logarithmic units. The most massive galaxies in our sample are about 0.44 dex below the main sequence, with 4 out of 5 less than 0.4 dex and one at 0.7 dex, consistently with our result that the quenching process is already under way in the cores of these galaxies.

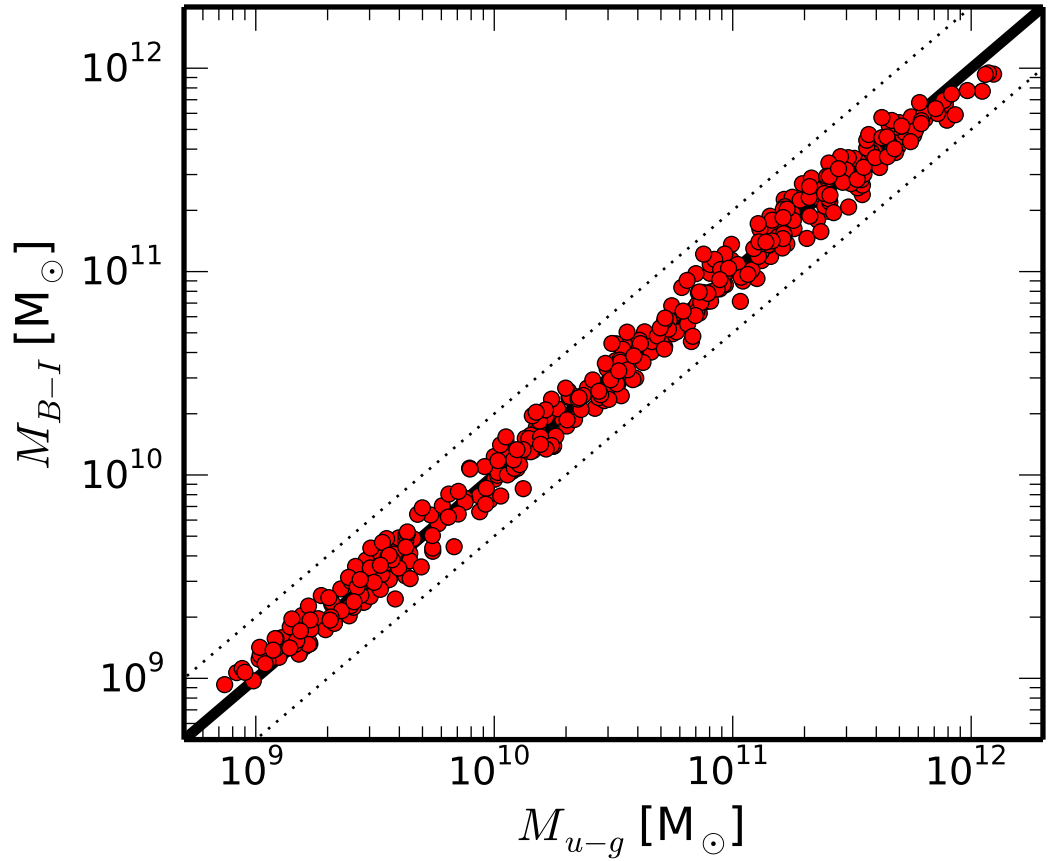

**Fig. S2. Testing stellar mass derivations.** We plot the comparison between the stellar masses obtained, for the SED model templates of our library, when using the rest-frame B-I color, as done for the ZENS galaxies, and the rest-frame u-g color, which is the input color that is provided by the WFC3 J-H color at the redshift of our high-z sample. The black solid lines show the one-to-one relation, while the dashed lines indicate the 0.3 dex difference.

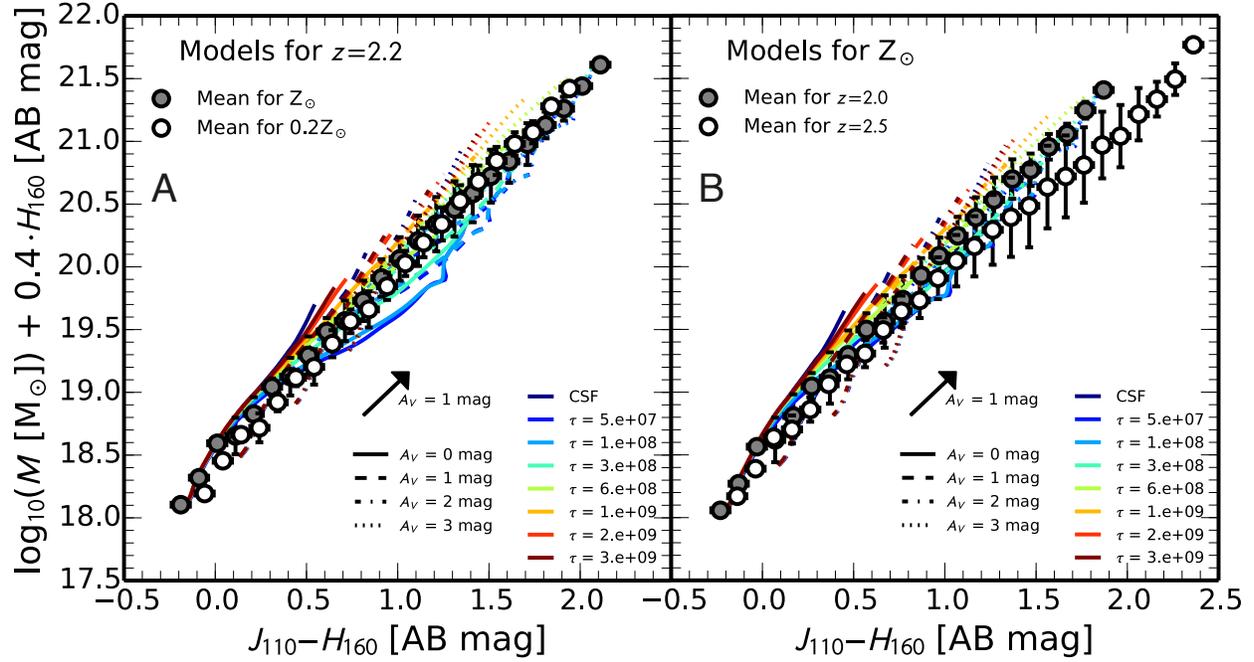

**Fig. S3. Color to mass-to-light ratio conversion.** We plot the relationship between observed $J_{110}$-$H_{160}$ colors and $\log_{10}(M)+0.4\cdot H_{160}$, which is proportional to the mass-to-light ratio. $H_{160}$ is the observed AB magnitude. The various colored curves are computed from Bruzual & Charlot models (*35*) with solar metallicity, a Chabrier IMF (*36*), and the Calzetti et al. reddening law (*37*). Different colors are used for different star formation histories: red to blue represent SFHs of increasing star formation timescale from a single stellar population. Different line styles are also used for models computed with different values for extinction. As shown by the arrow, the effects of extinction are roughly parallel to the locus of the various models. Panels (A) and (B) show the variation in metallicity and redshift, respectively. Such a relation was computed for each galaxy at its redshift and using the appropriate J-band filter.

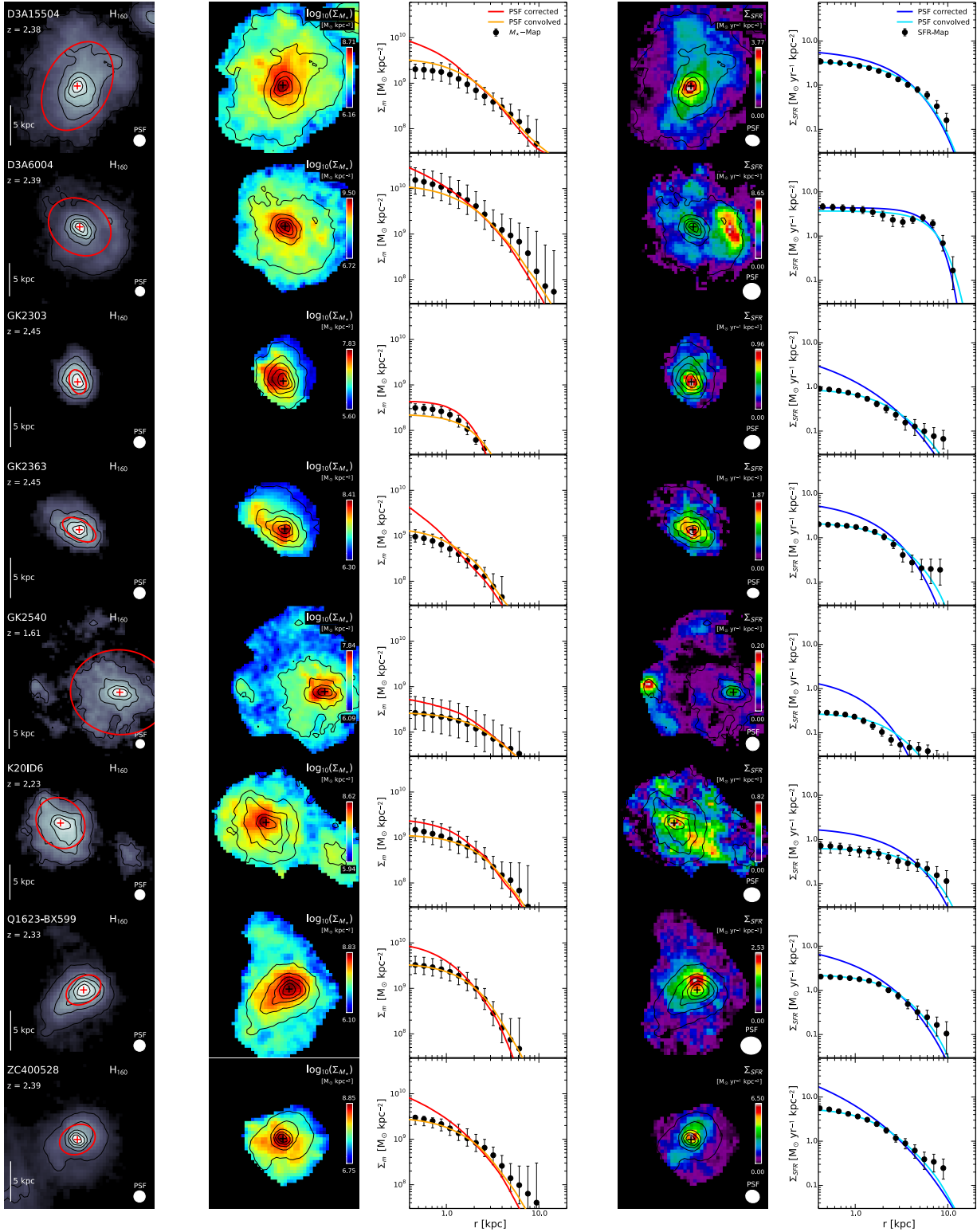

**Fig. S4. Galaxy by galaxy overview.** Each row represents a galaxy in our sample. The columns from left to right are: WFC3 H-band image, M map (in log scale), mass profile, SFR map, and SFR profile. The stellar mass profiles shown in red are the PSF-corrected profiles, and in orange

are the PSF-convolved profiles. The black data points are azimuthal averages measured from the $M_\star$ maps. Similarly, the blue and the cyan line show the PSF-corrected and PSF-convolved SFR profiles, respectively, while the black data points indicate the azimuthal averages measured from the SFR maps. The error bars are $1\sigma$ uncertainties arising from the combination of photometric errors and conversions to stellar mass and SFR.

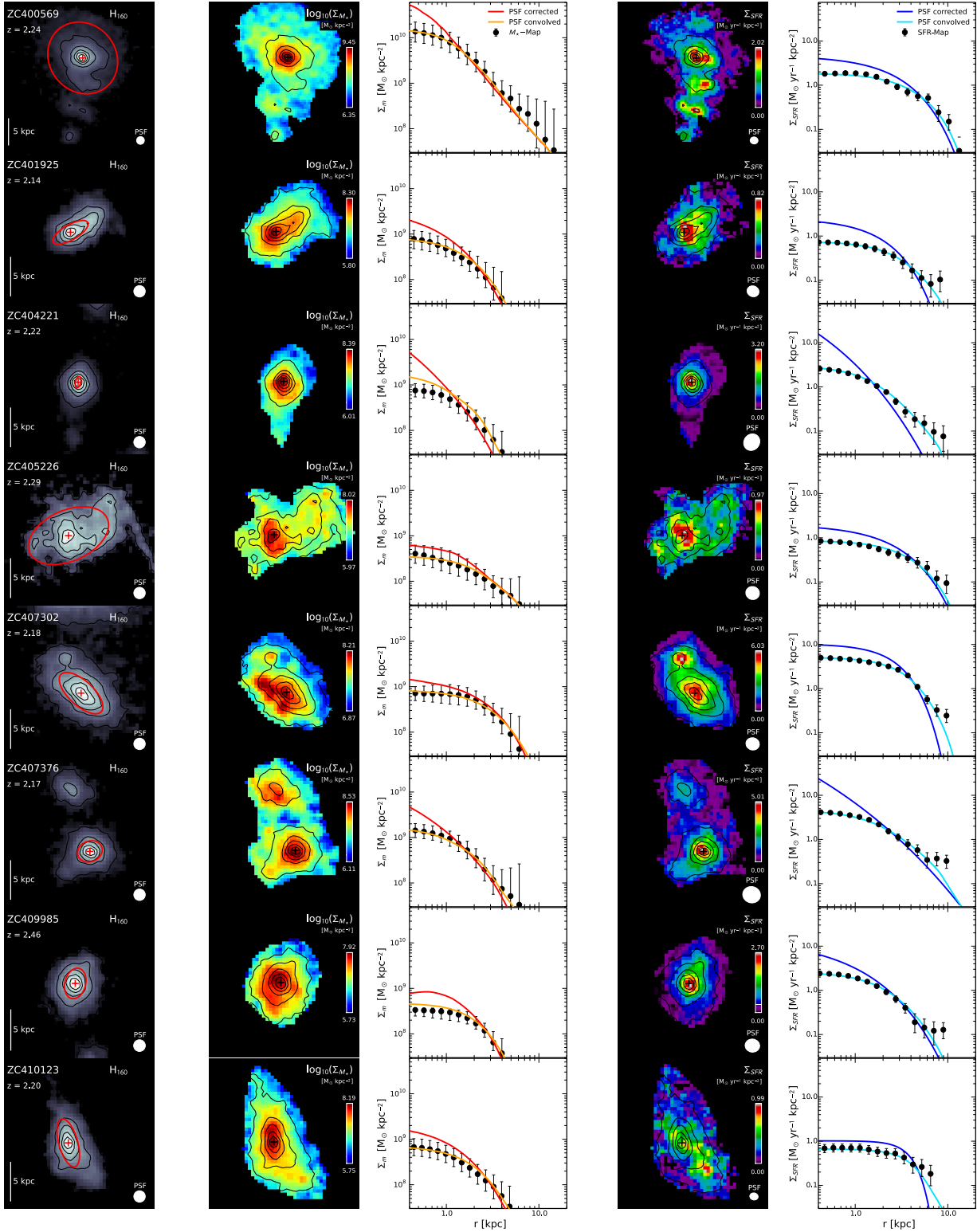

**Fig. S5.** Same as Fig. S4.

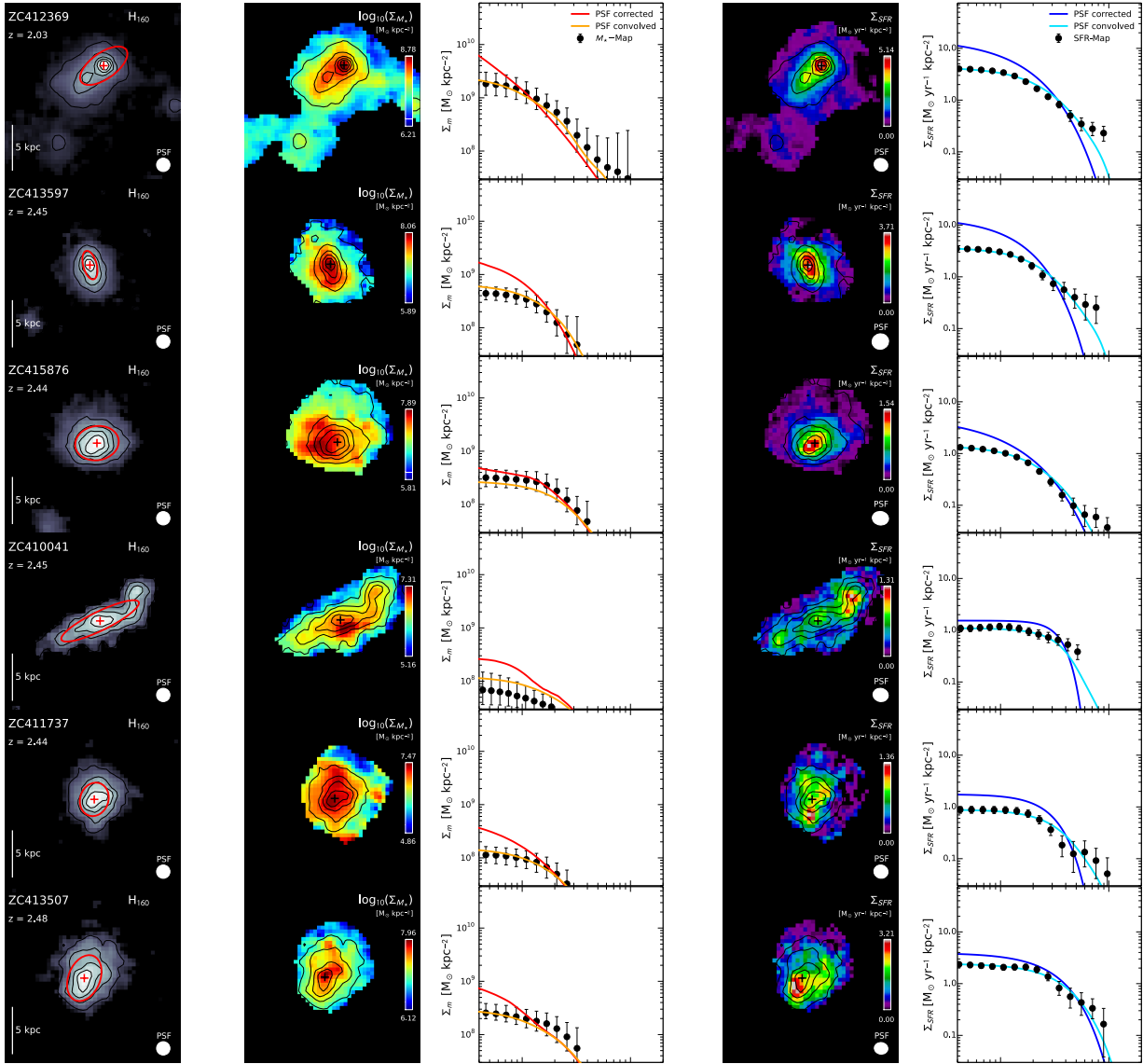

**Fig. S6.** Same as Fig. S4.

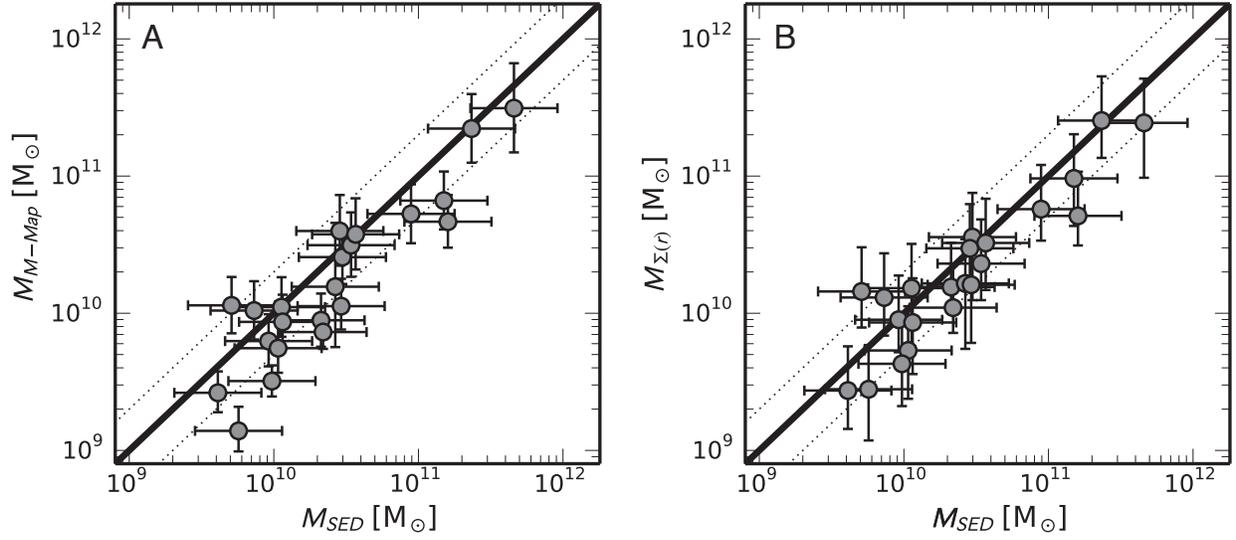

**Fig. S7. Verification of the stellar mass estimates.** We compare the integrated 2d mass maps $M_{M-Map}$ (A) and the integrated 1d profiles $M_{\Sigma(r)}$ (B) with the SED-derived stellar masses $M_{SED}$. The error bars indicate the systematic uncertainty in the derivation of stellar masses. The black solid lines show the one-to-one relation, while the dashed lines indicate the 0.3 dex difference. We find a good agreement between the integrated masses for all 22 galaxies in our sample.

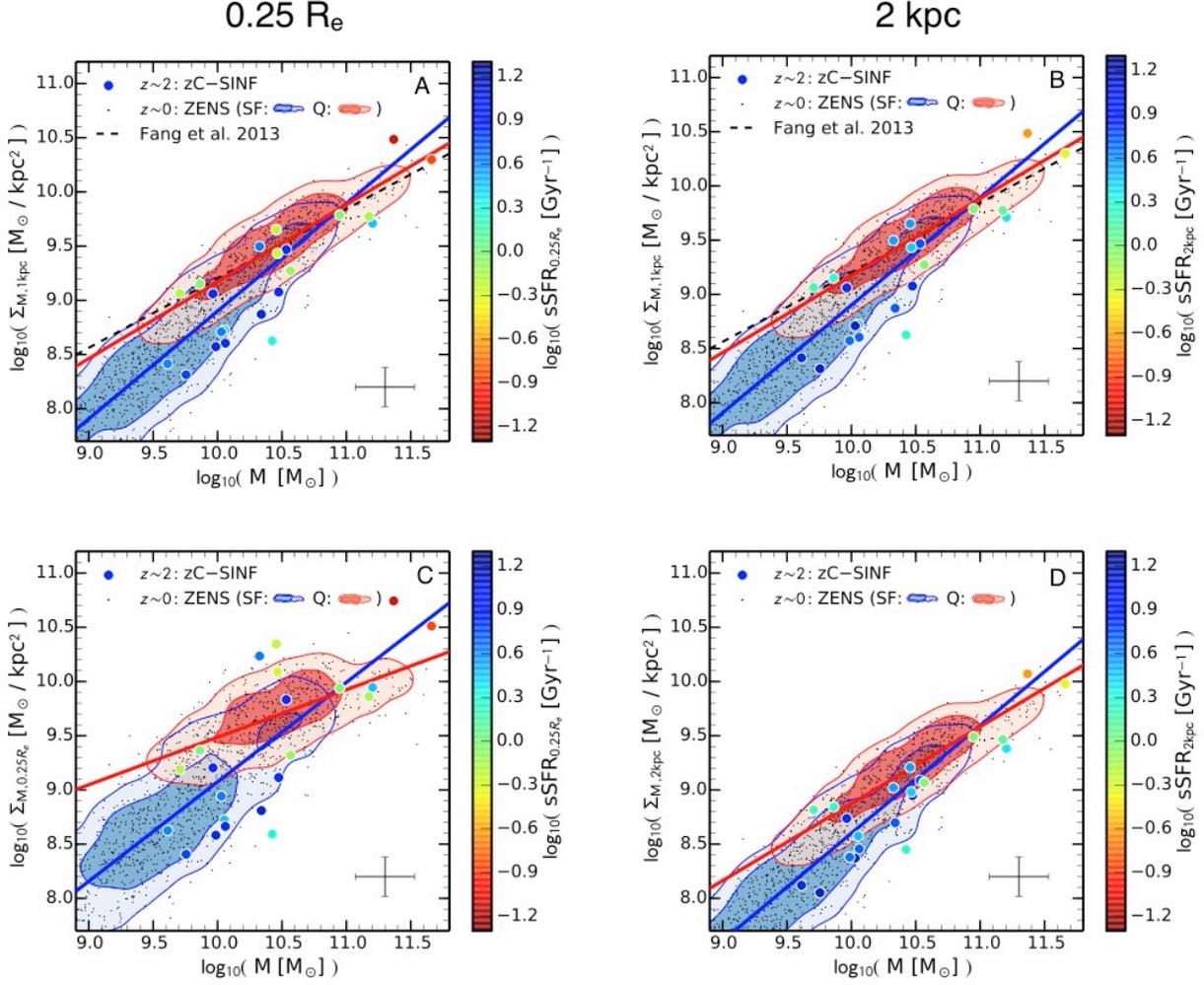

**Fig. S8. Central stellar mass density sequence with different apertures.** The panels shows the same as Fig. 2, but with two additional definitions of the central aperture. The upper panels show the central stellar mass density within 1 kpc as a function of the total stellar mass, where color coding corresponds to the sSFR within a quarter of the half-mass radius $R_e$, i.e. 0.25 $R_e$, (A) and within 2 kpc (B). The central star formation activity is reduced for the most massive systems regardless of the exact definition of the central sSFR. The lower panels reproduce similar relations but using the central stellar mass density and SFR density estimates matched in aperture within respectively 0.25 $R_e$ (C) and within 2 kpc (D).

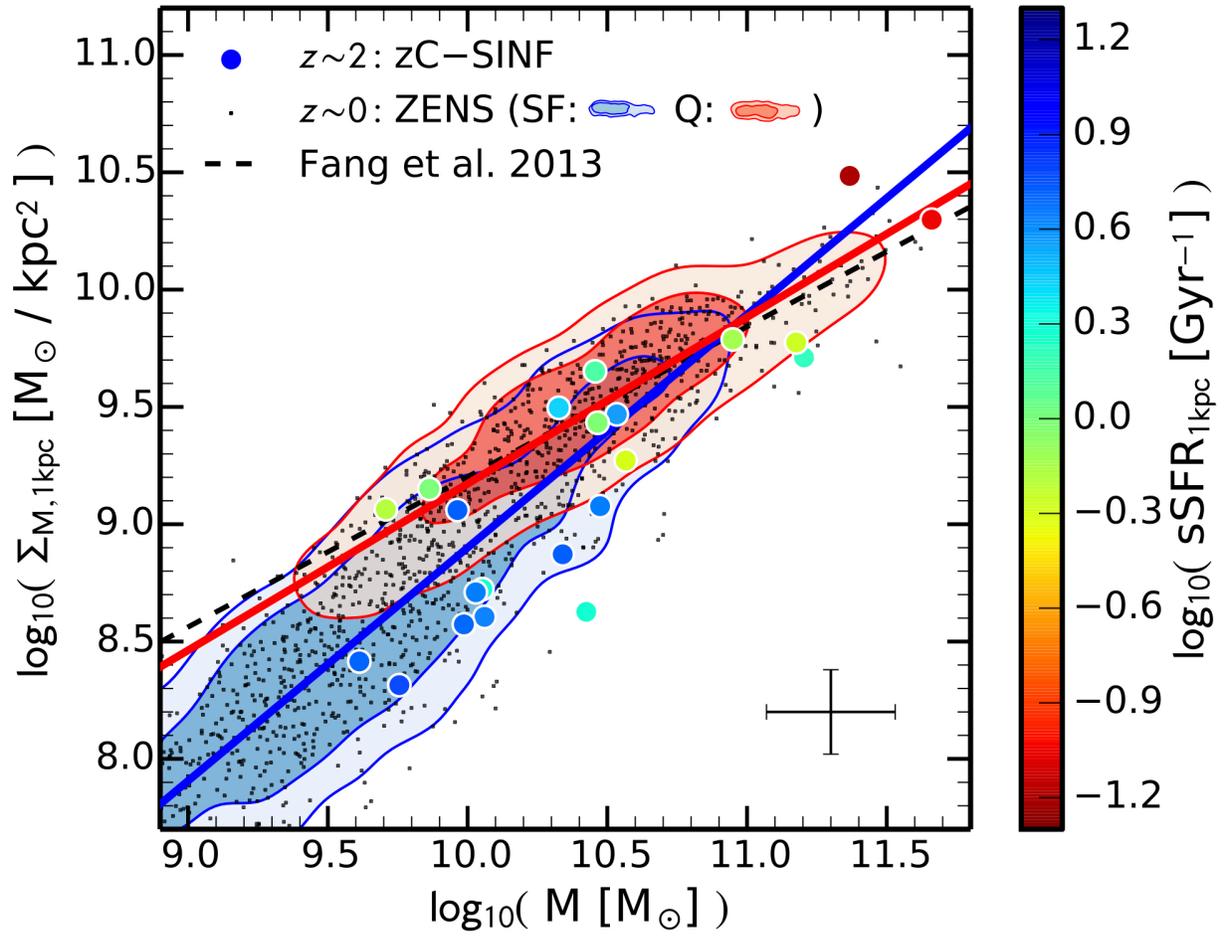

**Fig. S9. Central stellar mass density sequence with a different dust model.** Same as Fig. 2, but the sSFRs are computed using dust model Wuyts et al. (*26*). The reduction of the central sSFR is still observed.

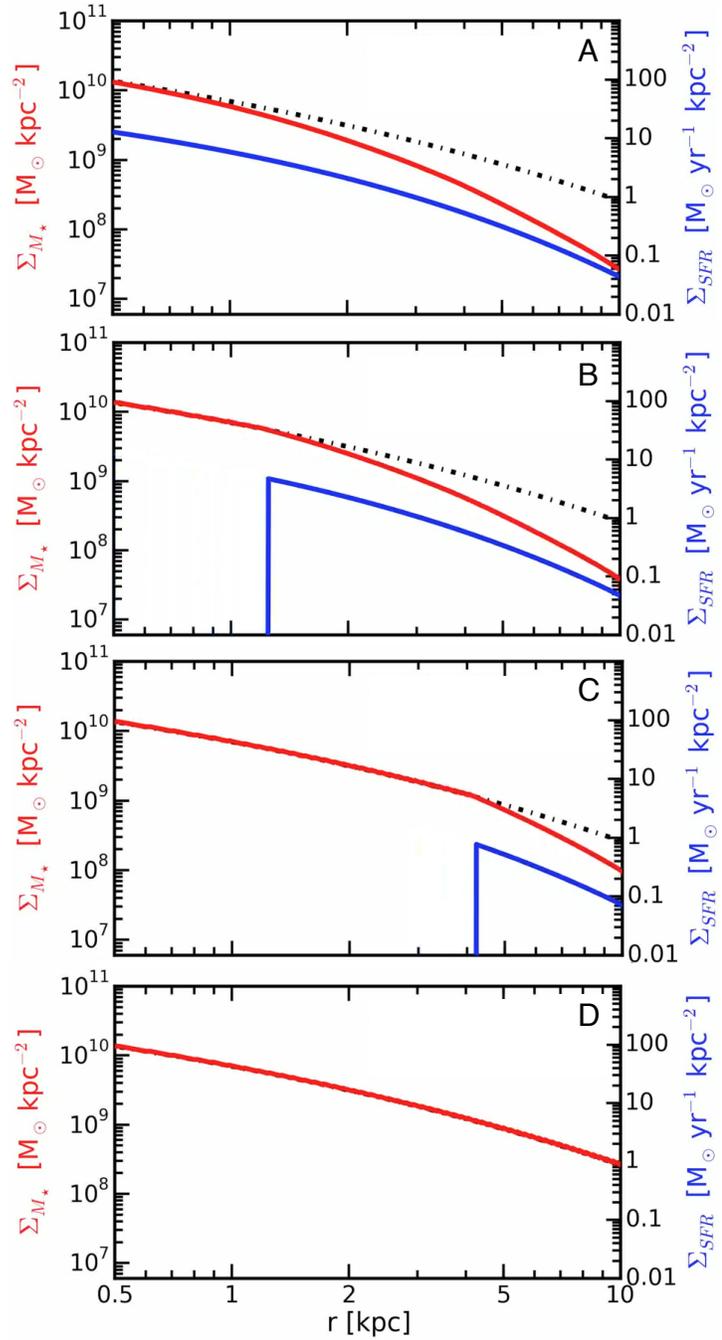

**Fig. S10. Results of our model.** The four panels (A-D) show snapshots ($z$ = 2.2, 2.1, 1.6, and 1.0) of the evolution of the mean profiles of the galaxies in our most massive bin. The mass profiles are shown in red, while the SFR profiles are shown in blue.

| Name | K [AB mag] | $z_{H\alpha}$ | M [$10^{10}$ M$_\odot$] | SFR [M$_\odot$/yr] | $A_{V,SED}$ [mag] |
|---|---|---|---|---|---|
| D3A-15504 | 21.28 | 2.3826 | 15.0 | 150 | 1.0 |
| D3A-6004 | 20.96 | 2.3867 | 45.8 | 210 | 1.8 |
| GK-2303 | 22.78 | 2.4501 | 0.97 | 21 | 0.4 |
| GK-2363 | 22.67 | 2.4518 | 2.92 | 64 | 1.2 |
| GK-2540 | 21.80 | 1.6146 | 2.66 | 21 | 0.6 |
| K20-ID6 | 22.14 | 2.2345 | 3.68 | 45 | 1.0 |
| Q1623-BX599 | 21.79 | 2.3313 | 8.89 | 34 | 0.4 |
| ZC400528 | 21.08 | 2.3876 | 16.0 | 148 | 0.9 |
| ZC400569 | 20.69 | 2.2405 | 23.3 | 241 | 1.4 |
| ZC401925 | 22.74 | 2.1411 | 0.73 | 47 | 0.7 |
| ZC404221 | 22.44 | 2.2201 | 2.12 | 61 | 0.7 |
| ZC405226 | 22.33 | 2.2872 | 1.13 | 117 | 1.0 |
| ZC407302 | 21.48 | 2.1814 | 2.98 | 340 | 1.3 |
| ZC407376 | 21.79 | 2.1733 | 3.42 | 89 | 1.2 |
| ZC409985 | 22.30 | 2.4577 | 2.19 | 51 | 0.6 |
| ZC410041 | 23.16 | 2.4539 | 0.57 | 47 | 0.6 |
| ZC410123 | 22.80 | 2.1987 | 0.51 | 59 | 0.8 |
| ZC411737 | 22.81 | 2.4443 | 0.41 | 48 | 0.6 |
| ZC412369 | 21.39 | 2.0283 | 2.86 | 94 | 1.0 |
| ZC413507 | 22.52 | 2.4794 | 1.07 | 111 | 1.1 |
| ZC413597 | 22.58 | 2.4498 | 0.92 | 84 | 1.0 |
| ZC415876 | 22.38 | 2.4362 | 1.15 | 94 | 1.0 |

**Table S1. List of galaxies in our sample.** Columns from left to right: Name, K-band AB magnitude, spectroscopic redshift from SINFONI, stellar mass, star-formation rate, and dust extinction from (*32*) and (*33*).